\documentclass[5p]{elsarticle}

\usepackage{lineno,hyperref}
\usepackage{dblfloatfix} 
\usepackage[table]{xcolor} 
\usepackage{xcolor} 
\usepackage{framed} 
\definecolor{shadecolor}{RGB}{10,10,10}
\usepackage{float}
\usepackage{graphicx}
\usepackage{amsmath}
\usepackage{caption}

\journal{ArXiv}

\begin{document}

\begin{frontmatter}

\title{3D HR-EBSD characterization of the plastic zone around crack tips in tungsten single crystals at the micron scale}

\author[mysecondaryaddress]{Szilvia Kal\'{a}cska\corref{mycorrespondingauthor}}
\cortext[mycorrespondingauthor]{Corresponding author}
\ead{szilvia.kalacska@empa.ch}
\author[mysecondaryaddress,myfirstaddress]{Johannes Ast}
\author[mythirdaddress]{P\'{e}ter Dus\'{a}n Isp\'{a}novity}
\author[mysecondaryaddress]{Johann Michler}
\author[mysecondaryaddress]{Xavier Maeder}

\address[mysecondaryaddress]{Empa, Swiss Federal Laboratories for Materials Science and Technology, Laboratory of Mechanics of Materials and Nanostructures, CH-3602 Thun, Feuerwerkerstrasse 39. Switzerland}
\address[myfirstaddress]{Fraunhofer IKTS, Fraunhofer Institute for Ceramic Technologies and Systems, Forcheim \"{A}u$\beta$ere N\"{u}rnberger Strasse 62, 91301 Forcheim, Germany}
\address[mythirdaddress]{E\"{o}tv\"{o}s Lor\'{a}nd University, Department of Materials Physics, 1117 Budapest, P\'{a}zm\'{a}ny P\'{e}ter s\'{e}tany 1/a. Hungary}

\begin{abstract}
High angular resolution electron backscatter diffraction (HR-EBSD) was coupled with focused ion beam (FIB) slicing to characterize the shape of the plastic zone in terms of geometrically necessary dislocations (GNDs) in W single crystal in 3 dimensions.
Cantilevers of similar size with a notch were fabricated by FIB and were deformed inside a scanning electron microscope at different temperatures (21$^{\circ}$C, 100$^{\circ}$C and 200$^{\circ}$C) just above the micro-scale brittle-to-ductile transition (BDT). J-integral testing was performed to analyse crack growth and determine the fracture toughness. At all three temperatures the plastic zone was found to be larger close to the free surface than inside the specimen, similar to macro-scale tension tests. However, at higher temperature, the 3D shape of the plastic zone changes from being localized in front of the crack tip to a butterfly-like distribution, shielding more efficiently the crack tip and inhibiting crack propagation. A comparison was made between two identically deformed samples, which were FIB-sliced from two different directions, to evaluate the reliability of the GND density estimation by HR-EBSD. The analysis of the distribution of the Nye tensor components was used to differentiate between the types of GNDs nucleated in the sample. The role of different types of dislocations in the plastic zone is discussed and we confirm earlier findings that the micro-scale BDT of W is mainly controlled by the nucleation of screw dislocations in front of the crack tip.

\end{abstract}


\end{frontmatter}


\section{Introduction}

Dislocation accumulation and distribution around crack tips play a great role in fracture mechanics. However until now, the characterization of the plastic zone around crack during fracture tests remains a difficult task. In recent years, a lot of progress has been made in understanding and quantifying fracture mechanics at small scales \cite{Pippan2018, Jaya2015, Ast2019}. The introduction of the J-integral technique at the micro-scale by Wurster et al. \cite{Wuster2012} has opened the possibility to analyse semi-brittle fracture processes at small scales and calculate the fracture toughness of materials showing non-negligible plastic deformation during fracture. Such behaviour is typical for body centered cubic (BCC) metals, which are usually found in numerous industrial applications, where their mechanisms of deformation need to be accurately defined for ambient and non-ambient conditions \cite{Rieth2013}. 

Local fracture behaviour of specifically oriented specimens of W single crystals was studied by Bohnert et al. \cite{Bohnert2016} by notched microcantilever tests, for which both numerical and experimental investigations were combined. Size-dependent fracture toughness of W was later studied by Ast \emph{et al.} \cite{Ast2017}. Their work particularly shows the importance of the plastic zone size at the crack tip on the fracture behaviour. At the macro-scale, the radius of the plastic zone ($r_{pl}$) in a plane strain--stress state in front of a crack tip has been estimated by Irwin \cite{Irwin1957}:

\begin{eqnarray}\label{eq:01}
r_{pl}=\frac{1}{3\pi}\left(\frac{K_I}{\sigma _y}\right)^2,
\end{eqnarray}
where $\sigma _y$ is the macroscopic yield stress and $K_I$ is the stress intensity factor, that can be expressed in terms of the stress and the crack length. Numerous works have been published to refine Irwin’s estimation \cite{Larsson1973, Betegon1991, Tracey1976, Kujawski1986, Vasco-Olmo2016}, but none of them can be applied to single crystals showing a strong anisotropy in plastic deformation or at small scale, where the size of the specimen approaches the size of the plastic zone. In fact, very few direct plastic zone characterization close to and in front of the crack tips have been done \cite{Murphy2009}. John’s work on the brittle-ductile transition (BDT) in pre-cleaved Si single crystal shows X-ray topographs of the crack-tip region where the size of the deformation-affected area can be recognized for different deformation conditions \cite{John1975}. A few TEM studies have also been conducted to understand the dislocation emission from the crack tip \cite{Horton1982, Ohr1984, Jagannadham1998}. 

High angular resolution electron backscatter diffraction (HR-EBSD) during \emph{in situ} notched microcantilever testing has been used to map the evolution of geometrically necessary dislocations (GNDs) at the crack tip at progressive deformation steps in W single crystal \cite{Ast22017, Ast22018}. \emph{Post mortem} HR-EBSD has been also applied to study the distribution of the plastic zone in front of the crack tip of W single crystal just above the BDT for different strain rates and temperatures (-90$^{\circ}$C to 500$^{\circ}$C) \cite{Ast2018}. This work revealed that the BDT at the micron-scale occurs at significantly lower temperature than the BDT in macro-scale testing \cite{Giannattasio2008, Gumbsch1998}. This difference is due to the size and shape of the plastic zone. At small scales and comparably low strain rates, in the BDT regime, the mobility of dislocations is already sufficiently high at lower temperatures because the activation energy for kink-pair formation for the screw dislocations is reduced \cite{Seeger1981, Brunner2000}. Therefore, dislocations emitted from the crack tip can escape fast enough through the free surfaces and further dislocations can be nucleated. At higher temperature, in the plastic regime, the crack tip is completely shielded by a higher dislocation density in front of the crack tip, having a higher ratio of screw components.

Another interesting aspect of mechanical testing of micron sized cantilevers is to study the free surface--dislocation interaction. 3-dimensional finite element (FE) framework \cite{Husser2017} and discrete dislocation dynamics simulations \cite{Motz2008, Tarleton2009, Tarleton2015} were applied to study dislocation structures during cantilever beam bending experiments. In reality, the presence of free surface modifies the stress gradient for dislocation motion, therefore, at the micro-scale stresses are sufficiently different from the bulk stresses \cite{Hazzledine1982}. To investigate this phenomenon, a surface-sensitive technique must be applied.

HR-EBSD can characterize the distribution of GNDs with a sub-100 nm step size, it is however limited by the fact that information is only gathered from the first few tens of nanometers below the surface \cite{Chen2011}. The 3D shape of the plastic zone at the crack tip in compact tension specimens is known to have a “dog bone” shape, being smaller inside the volume than towards the surfaces. This is due to the fact that plane strain occurs in the centre of the specimen, while plane stress state occurs at the surface \cite{Ewalds1984}. The true shape of the plastic zone in front of the crack tip remains however largely unknown at the micron-scale, where the free surface-effect is more pronounced. Recently, it was demonstrated on compressed copper single crystal micropillars \cite{Kalacska2020} and copper bicrystal cantilevers \cite{Konijnenberg2015} that 3-dimensional HR-EBSD can be successfully applied to characterize GND distribution in the entire volume of deformed samples.

In the present work, we apply the novel 3D HR-EBSD technique to characterize the 3-dimensional GND distribution in front of the crack tip of notched microcantilevers in body-centered cubic W, deformed at different temperatures and strain rates. Microcantilever beam bending is chosen as form of deformation because many of the formed dislocations during the plastic deformation will accommodate the bending of the crystal and can then be easily characterized by HR-EBSD in terms of GNDs. Our aim is to discuss differences in GND density distributions caused by elevated temperatures and to study the distribution of edge and screw dislocations in order to further elucidate the mechanisms leading to a shift in BDT to lower temperatures in this size regime \cite{Ast2018}. Note that brittle or semi-brittle fracture leads to loss of specimens at this length scale and makes subsequent HR-EBSD cha\-rac\-te\-ri\-za\-tion impossible. For this reason, the range starting from room temperate was explicitly picked. By reducing the loading rates cleavage could be suppressed so that the respective specimens could still be characterized by HR-EBSD after mechanical testing and to study dis\-lo\-ca\-tion-\-me\-di\-a\-ted stable crack growth in the specimens in 3D.

\section{Experimental methods}

\subsection{Sample preparation}

\begin{figure}[!ht]
\begin{center}
\includegraphics[width=0.45\textwidth]{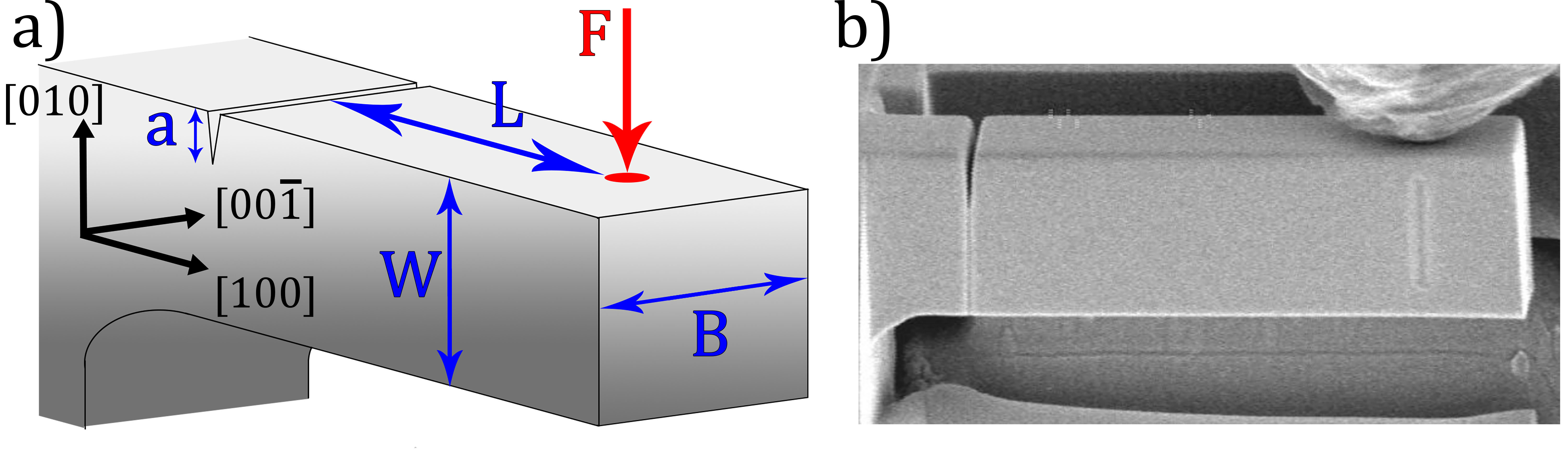} \\
\caption{\label{fig:01} a) Sketch of the prepared microcantilevers. Denotations: {\bf a} -- notch depth, {\bf B} -- cantilever thickness, {\bf L} -- distance between the bending point and the notch (lever arm), {\bf W} -- width of cantilever. Crystal orientation of the three main surfaces are presented by black arrows. b) Secondary electron image of a cantilever during bending. FIB marker on the side helped in the accurate positioning of the conical tip.}
\end{center}
\end{figure}

A high-purity W single crystalline sample with a (100)[001] type crack system were studied, where (100) is the crack plane, and $[001]$ is the direction of crack front propagation. In W, the primary cleavage planes are the \{100\} type planes \cite{Dull1965}. The bulk sample was mechanically polished with SiC paper up to P4000 grit. In order to be able to analyse the cantilevers by HR-EBSD after deformation, it was necessary to place them on the edge of the bulk sample. Therefore, Ar ion milling was applied on two perpendicular surfaces by a Leica EM TIC 3X polisher at 10 kV to create a sharp edge. 9 identical cantilevers were prepared by a Tescan Lyra3 Ga$^+$ FIB with beam currents 30 kV, 10 nA decreasing to 30 kV, 0.5 nA to minimize the FIB-affected surface layer. The lower edges of the cantilevers were shaped with a rounding radius of 1.5 $\mu$m at the base to avoid stress concentration. The notch was milled using a line cut with 30 kV 230 pA from the top, and a FIB milled line marker shown in the scanning electron (SE) image in Figure \ref{fig:01} b) was positioned at $L \sim$14 $\mu$m from the notch on the side with beam settings of 30 kV 70 pA to obtain similar lever arms. All samples had an approximate length to width to height ratio of 14.5 $\mu$m/4.5 $\mu$m/5.4 $\mu$m. The denotations of the prepared microcantilevers can be seen in Figure \ref{fig:01} a).

\subsection{Microcantilever testing and evaluation}

In situ microcantilever tests were performed with a nanoindentation platform (Alemnis AG, Switzerland) equ\-ipped with a cono-spherical diamond indenter tip with a radius of ca. $1-2$ $\mu$m. Room temperature tests were carried out in a Philips XL-30 type scanning electron microscope (SEM), while high temperature measurements were performed in a Zeiss DSM962 SEM. To monitor and adjust the temperature locally, the sample and the tip holder were attached separately to thermal sensors and resistive heaters \cite{Wheeler2012}. Before mechanical testing, the temperature of the sample was balanced with the indenter tip temperature to minimize thermal drift upon contact. Thermal drift rates were below 10 nm/min at room temperature and below 20 nm/min at elevated temperatures during the displacement-controlled tests. During micromechanical testing the load-displacement data were recorded.

If not fractured earlier, comparable deformation levels were chosen in each test. The data were normalized for better comparison using the following equation taken from the ASTM standard E399 \cite{ASTM399}:

\begin{eqnarray}\label{eq:02}
K_I=\frac{FL}{BW^{3/2}} \times f\left( \frac{a}{W}\right),
\end{eqnarray}
where $f$ is the geometry factor. Some details of this geometry function can be found in \emph{Appendix A} along with further equations to describe the small scale fracture mechanical behaviour by means of J-Integrals.

By superimposing a sinusoidal displacement with an amplitude of 10 nm to the applied monotonic loading protocol, a contact stiffness measurement was performed during the tests. The procedure is explained in more detail in previous studies \cite{Ast2016, Ast2019}. This was done to follow crack growth in the samples. All specimen showed non-negligible crack tip plasticity, while elastic-plastic fracture mechanics (EPFM) were used to determine the fracture toughness characterized by the onset of stable crack growth, which is denominated as "fracture initiation toughness".

From the obtained R-curves critical J-Integrals ($J_q$) are determined according to different criteria for the onset of stable crack propagation. This is shown in more detail in the Results chapter. Using the following equation, the conditional fracture toughness is calculated as:

\begin{eqnarray}\label{eq:02final}
\begin{aligned}
K_{Iq,J}=\sqrt{\frac{J_{q}E}{(1-\nu^2)}},
\end{aligned}
\end{eqnarray}
where $E$ is the Young's modulus, $\nu$ is the Poisson's ratio, and $q$ marks the application of a critical force in the fracture toughness calculation.

\subsection{3-dimensional HR-EBSD measurements}\label{sec:3D}

For the 3-dimensional HR-EBSD measurements, a Tescan Lyra3 dual beam SEM equipped with an Edax Digiview EBSD camera was used. The chamber of this SEM is specially designed to allow rotation-free (static) 3D EBSD measurements, reducing any drift originating from stage movements. The samples were placed at a 70$^\circ$ pre-tilted position, then consecutive FIB slicing and EBSD mapping was performed at a working distance of 9 mm. Slicing was carried out with beam conditions of 30 kV, 450 pA, while the slicing thickness was 90 nm. EBSD mapping was done using 20 kV, 16 nA beam with 100 nm step size on a square grid. The applied 2$\times$2 binning with zero gain level enabled collection of diffraction patterns with 442 px $\times$ 442 px resolution.

HR-EBSD evaluation was done using CrossCourt v4, a cross-correlation based analysis tool  \cite{Wilkinson2006} to calculate strain and stress tensor elements. GND densities were estimated from the lattice distortion tensor ($\beta_{ij}$) and using the approach of minimizing the strain energy ratio between screw and edge dislocations \cite{Wilkinson2010}, namely $E_{edge}/E_{screw}=1/(1-\nu)$ that leads to a lower bound estimation of the GND density. 24 slip systems were considered for the analysis of W, both \{110\} and \{112\} type slip planes and $<111>$ type slip directions, which can be divided into 24 edge and 8 screw types of dislocations. To complete this so called $L^1$ optimization scheme \cite{Wilkinson2010}, the $\alpha_{i3}$ Nye dislocation tensor components from the $\beta_{ij}$ deformation gradient tensor elements were calculated by a self-written $C++$ code. Through $\alpha_{i3}$ components, edge type ($\alpha_{13}$, $\alpha_{23}$) and screw type ($\alpha_{33}$) GNDs can be distinguished, therefore their evolution can be studied separately. Only these three Nye tensor components can be calculated without any assumption, hence they were used to study the distribution of edge and screw components (not relying on the $L_1$ optimisation method). Furthermore  $\alpha_{sq}=\sqrt{\alpha_{13}^2 + \alpha_{23}^2 + \alpha_{33}^2}$ values were determined, where $\alpha_{sq}$ is proportional to the GND density but it has the advantage to rely only on the straightforward calculation of $\alpha_{i3}$ components \cite{Kalacska2020}.

In this article, we utilize both GND estimation through $L_1$ optimisation and $\alpha_{sq}$ calculation technique on the same dataset in order to study GND density distribution and estimate their total density present in the system. 

In order to examine the whole volume of the plastic zone, one cantilever (named S3) was lifted out using a nanomanipulator, then re-attached to a copper TEM lamella holder, where FIB slicing and HR-EBSD mapping was performed. A thin Pt layer was deposited on top of the cantilever to reduce FIB damage and protect the integrity of the sample. FIB slicing and EBSD mapping were done on this sample identically to the other measurements in this study, only the FIB slicing direction was perpendicular to the side cutting geometry (see sketches in Fig. \ref{fig:06} later on). One diffraction pattern from the fixed end of the cantilever remote to any plastic activity was chosen as reference to make the evaluation unified. In case of this sample, all HR-EBSD calculated values relate to this single reference, which was assumed to be at minimal stress state. The plastic zone in cantilevers can only be totally mapped in 3D with the lift-out geometry, because in the side cutting case as the sample gets thinner by the FIB slicing, eventually it will bend under its own weight or from residual stresses once it gets thin enough. This would introduce artefacts in the HR-EBSD maps that are not part of the initial plastic zone.

\section{Results}

\subsection{Fracture behaviour}

\begin{figure*}[!ht]
\begin{center}
\includegraphics[width=0.8\textwidth]{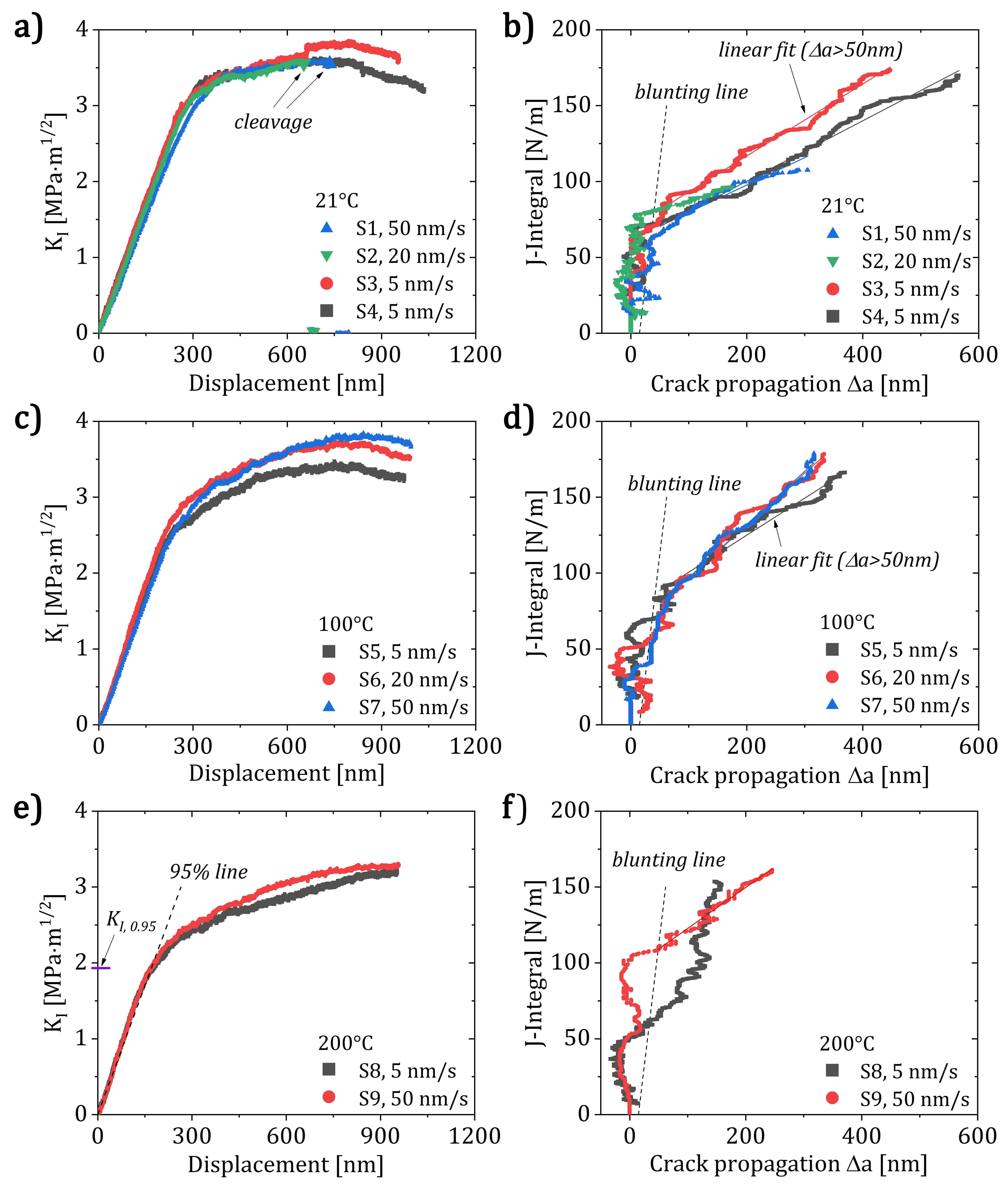} \\
\caption{\label{fig:02} a), c) and e) show plots of the stress intensity factor ($K_I$) as a function of bending displacement for specimens tested at room temperature (21$^{\circ}$C), 100$^{\circ}$C and 200$^{\circ}$C, respectively. Specimens were deformed at different displacement rates ($5-50$ nm/s). In e) a dashed line labelled "95\% line" corresponds to a construction line, that pass through the origin and have 95\% of the slope of the linear fits of the elastic part of the dataset. The intersection of this construction line and the mechanical data determined $K_{IQ,LEFM}$ (blue marker on the y axis). b), d) and f) correspond to respective J-Integral -- crack propagation plots. They describe how the fracture initiation toughness is determined by means of the intercept of a blunting (dashed) line and a linear fit (continuous line) of data (for $\Delta a >50$ nm).}
\end{center}
\end{figure*}

\begin{table*}[!ht] 
\begin{tabular}[width=\textwidth]{|c|c|c|c|c|c|c|l|}\hline 
Name & Temp. & Loading& $K_{Iq, LEFM}$  &  \multicolumn{3}{c|}{$K_{Iq, J}$  [MPa$\cdot$m$^{1/2}$] } & Remarks\\  \cline{5-7}

 & [$^{\circ}$C] & rate [nm/s] & [MPa$\cdot$m$^{1/2}$] & $\Delta a = 0$ nm & Blunting line  & $\Delta a = 200$ nm & \\
\hline 
S1 & 21 & 50 & 3.0 & 5.2 & 5.5 & 6.7 & Cleavage \\
S2 & 21 & 20 & 3.1 & 5.8 & 6.0 & - & Cleavage \\
S3  & 21 & 5 & 3.0 & 5.6 & 5.9 & 7.3 & Lift-out 3D  \\
S4 & 21 & 5 & 3.1 & 5.2 & 5.5 & 6.5 & 3D  \\
S5 & 100 & 50 & 2.5 & 5.8 & 6.2 & 7.5 &- \\
S6 & 100 & 20 & 2.6 & 5.3 & 5.8 & 7.9 & - \\
S7 & 100 & 5 & 2.6 & 5.2 & 5.8 & 7.6 & 3D \\
S8 & 200 & 50 & 1.9 & - & - & - & -\\
S9 & 200 & 5 & 1.9 & 6.4 & 6.9 & 8.1 & 3D \\
\hline

\end{tabular} 
\centering \caption{Overview of tested specimens and fracture toughness according to LEFM by means of the 95\%  line and EPFM by means of the J-Integral technique; for the latter three different approaches were used to determine the onset of fracture, the fracture initiation toughness.} 
\label{table:01} 
\end{table*}
 
Different displacement rates (5 nm/s, 20 nm/s and 50 nm/s) were applied to different samples to investigate the previously observed effect of loading rate and temperature on small-scale fracture behaviour. The goal was to better understand the micromechanical response in these specimens that can be related to the development of the plastic zones. The given  displacement rates, which correspond to stress intensity factor rates of approximately 0.06 MPa$\cdot$m$^{1/2}s^{-1}$, 0.24 MPa$\cdot$m$^{1/2}s^{-1}$ and 0.6 MPa$\cdot$m$^{1/2}s^{-1}$, respectively, and temperatures were chosen such that the micro-scale BDT occurs in this range.

Initially, linear-elastic loading behaviour is observed for all specimens and all investigated temperatures, as shown in  Figure \ref{fig:02} a), c) and e). Since the transition from elastic to elasto-plastic loading is smooth, a criterion from ASTM standard E399 \cite{ASTM399} is chosen to define the end of elastic loading, which is at the same time the lower bound of the fracture toughness as described by LEFM. This lower bound value was determined by performing linear fits of the linear-elastic data. Then construction lines were introduced, which have 95 \% of the respective slopes of the linear fits and which pass through the origin. The intercept of these construction lines and the mechanical data yields $K_{Iq, LEFM}$ and the procedure is exemplarily shown in Figure \ref{fig:02} e). Irrespective of the applied displacement rate, the tests at 21$^{\circ}$C show $K_{Iq, LEFM} = (3.0-3.1)$ MPa$\cdot$m$^{1/2}$, also shown in Table \ref{table:01}. With increasing temperature, the onset of yielding decreases and the conditional fracture toughness according to LEFM drops to $\sim$ 2.6 MPa$\cdot$m$^{1/2}$ at 100$^{\circ}$C and 1.9 MPa$\cdot$m$^{1/2}$ at 200$^{\circ}$C and remain independent of the loading rate. However, these values do not describe the fracture behaviour and are only given as a matter of completeness. At 100$^{\circ}$C and particularly at 200$^{\circ}$C significant crack tip plasticity is observed, which is characterized by the type of hardening behaviour.

At 21$^{\circ}$C, tests using displacement rates of 50 nm/s and 20 nm/s resulted in cleavage fracture, as shown in Figure \ref{fig:02} a). At 5 nm/s more stable crack propagation was achieved, cleavage was suppressed and tests were manually stopped at total bending displacements of approximately 1000 nm. This BDT in micron-sized specimens made from high purity W single crystals and for the investigated \{100\}$<$100$>$ crack system was studied in detail in ref. \cite{Ast2018}. Using the same displacement rates, the other two investigated temperatures showed only micro-cleavage behaviour, which is characterized by stable crack growth.

EPFM was applied for the determination of fracture toughness from crack resistance curves, which are shown in Figure \ref{fig:02} b), d) and f) for the different temperatures. Three different approaches were chosen to describe the fracture initiation toughness. The first one, being the most conservative one, aims at calculating conditional J-Integrals, which characterize the very onset of fracture ($\Delta a = 0$ nm) by linear extrapolation of the R-curve data for $\Delta a > 50$ nm. This criterion was chosen to exclude data from the blunting regime and works well for data recorded at 21$^{\circ}$C and 100$^{\circ}$C. For the second technique the same linear fit is applied but this time a blunting line is used and the intercept of this blunting line and the linear fits of the R-curve data yield the conditional J-Integrals. The third method, already described in ref. \cite{Ast2016} uses the intercept of the fitted R-curve data and a line at $\Delta a > 200$ nm. Since enhanced crack tip plasticity and very limited crack growth took place at 200$^{\circ}$C, only one sample (S9) could be successfully evaluated according to these criteria. An overview of all fracture toughness data, which were calculated from the conditional J-Integrals according to Eq. (\ref{eq:02final}) are also shown in Table \ref{table:01}. With increasing temperature the R-curve behaviour changes and J-Integrals rise more significantly once crack growth occurs. This shows that the resistance to crack propagation increases with increasing temperature as expected.

\begin{figure}[!ht]
\begin{center}
\includegraphics[width=0.45\textwidth]{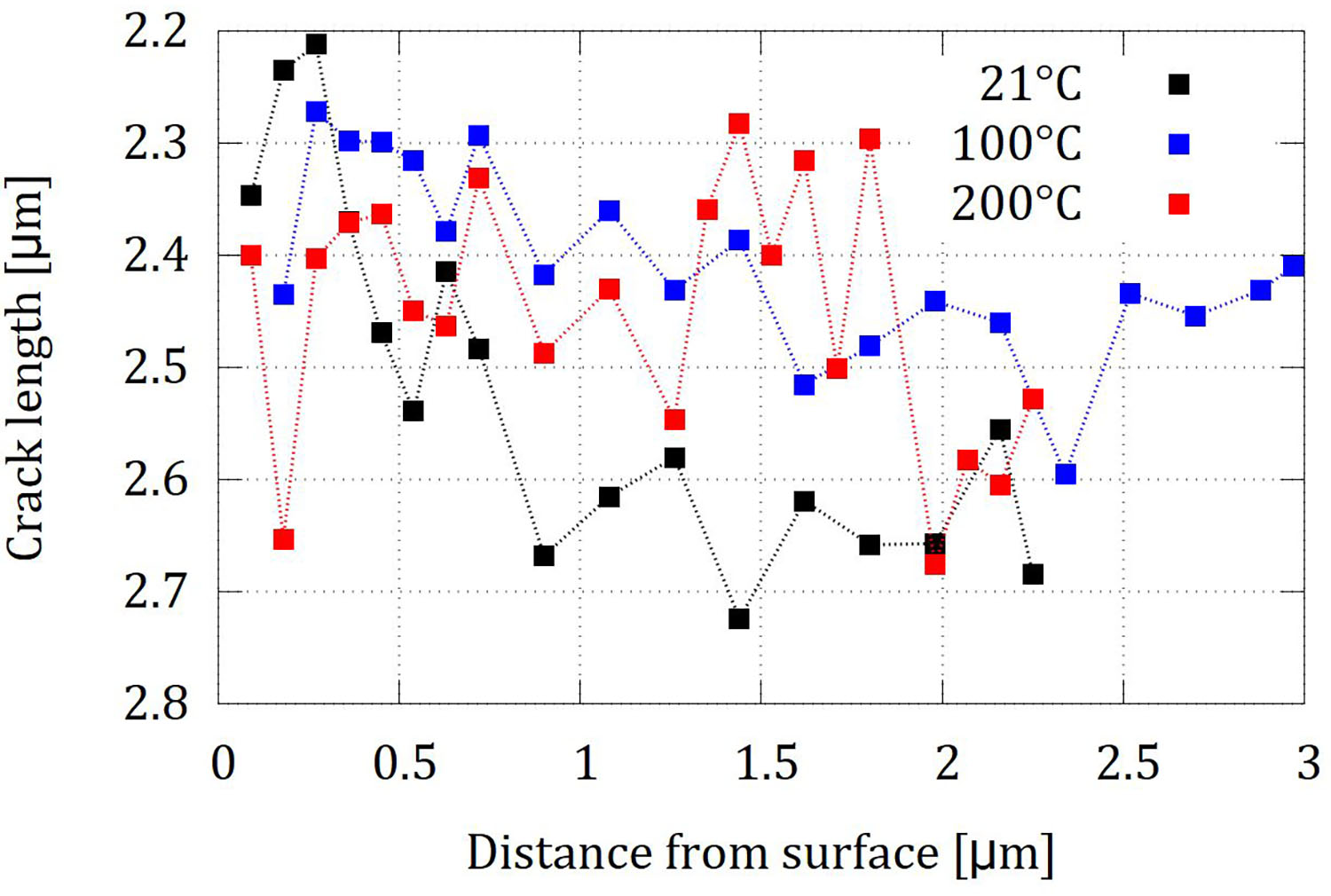} \\
\caption{\label{fig:R2} Plot of the crack length as a function of distance from the surface for 3D reconstructed specimens (S4 - 21$^{\circ}$C, S7 - 100$^{\circ}$C and S9 - 200$^{\circ}$C).}
\end{center}
\end{figure}

Unfortunately, it was not possible to deform the samples to a specific strain field or up to a certain GND arrangement. The only way to compare specimens of similar size and shape was to perform tests to similar bending displacements and to analyze respective changes in the evolution of the GND arrangements. In order to try to assess the crack length at different temperatures, a crack-extension me\-asu\-re\-ment was conducted on the secondary electron images recorded on the 3D reconstructed samples before each HR-EBSD measurement. Although the images were not high resolution, the evaluation strategy was able to conclude crack lengths on all three samples as a function of distance from the surface (Figure \ref{fig:R2}). The data is provided as supp\-le\-men\-tary material, where the description of the evaluation can be found.

Usually crack extension is investigated on cleaved sur\-fa\-ces, but in this case that was not possible. However, a simple observation could be made on the crack shape. At room temperature the crack seems to be deeper in the middle of the cantilever than closer to the free surface. Also from the sample deformed at 200$^{\circ}$C (S9, which was sliced up beyond its half volume) this tendency seems to be symmetric along the crack tip, as values of the crack length begin to decrease again beyond 2.5 $\mu$m distance from the surface. An attempt to carry out similar evaluation on the lift-out sample was also made, but due to FIB milling artefacts a clear image of the notch was not possible to find.

\subsection{3D HR-EBSD analysis}\label{sec:3D2}

\begin{figure*}[!ht]
\begin{center}
\includegraphics[width=0.95\textwidth]{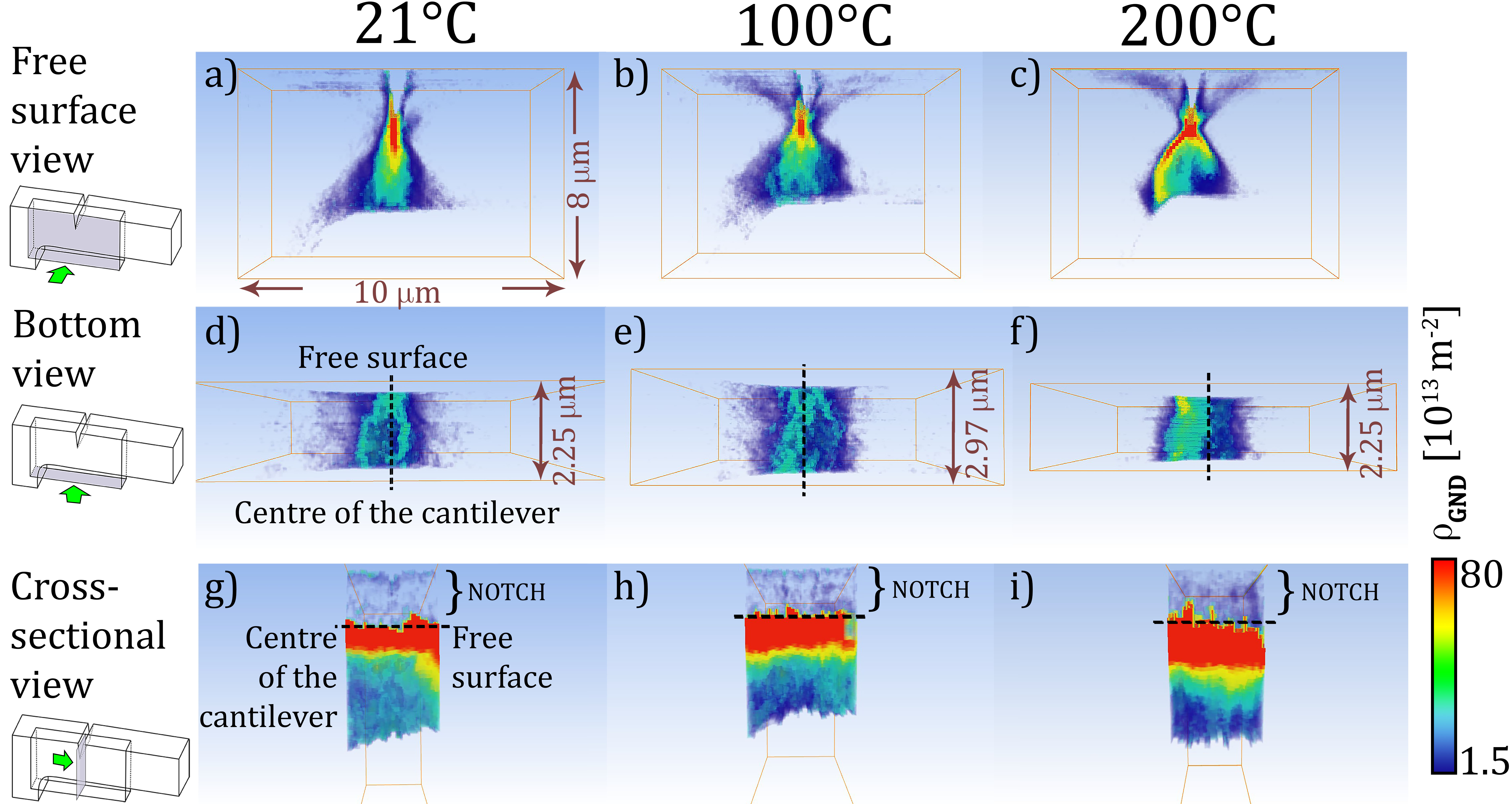} \\
\caption{\label{fig:03} Total GND density reconstructions calculated for cantilevers deformed at three different temperatures (21$^{\circ}$C, 100$^{\circ}$C and 200$^{\circ}$C with 5 nm/s tip velocity), viewed from the free surface (a, b, c), from the bottom (d, e, f) and the cross-section (g, h, i) of the sample at the crack position. Dashed lines indicate the position of the crack front. 3D mapping was carried out only on the half volume of each cantilever. Sketches on the side in each row show with grey areas from where the reconstruction is viewed. Supplementary videos are provided of the three reconstructions.}
\end{center}
\end{figure*}

Figure  \ref{fig:03} shows the total 3D GND density ($\rho_{GND}$) reconstruction of the three cantilevers deformed at 21$^{\circ}$C, 100$^{\circ}$C and 200$^{\circ}$C, with various views. The rotational views of these reconstructions as well as the plastic zone shapes can be studied in the supplementary videos in more detail. Note that these reconstructions created from the side cutting geometry only show half of the total volume.
Viewed from the side (Figure \ref{fig:03} a, b, c) the lateral size of the plastic zone does not increase significantly with temperature, but the amount of GNDs in front of the crack tip progressively does, with also more backward dislocation generation towards the top of the cantilever, giving a characteristic "dog bone" shape of the plastic zone around the crack tip at elevated temperatures. These are similar observations to those that are reported in a previous work \cite{Ast2018}.

The 3D shapes of the plastic zones of these three deformation conditions can be studied by plotting various cross-sections. Figure \ref{fig:03} d), e), and f) shows the total GND distribution viewed from the bottom, right in front of the crack tip, perpendicular to the free surface. This view shows that the high GND concentration zone appears symmetrically distributed in front of the crack tip in the cantilever that was deformed at 21$^{\circ}$C (Fig.\ref{fig:03} d). This high GND density zone is below the projected crack front near the free surface, and on both sides of the crack front towards the centre of the specimen, following the projection of the $\{110\}$ planes. At 100$^{\circ}$C (Fig.\ref{fig:03} e) the GND distribution appears to be less symmetric around the crack front, which can be due to the activation of other slip systems like $\{112\}$ and $\{123\}$. The specimen deformed at 200$^{\circ}$C (Fig.\ref{fig:03} f) shows much more GNDs, preferentially concentrated on one side towards the support of the cantilever.

Looking at the cross-sectional distribution in front of the crack tip for the cantilever deformed at 21$^{\circ}$C (Fig.\ref{fig:03} g), one can see more GNDs towards the free surface until a depth of about 1 $\mu$m, giving a "dog bone" shape of the plastic zone. This is similar to what is presented in Figure \ref{fig:04} for the lift-out sample later on. This "dog bone" shape seems to disappear at higher temperatures (Fig. \ref{fig:03} h  and i), where the plastic zone is more homogeneously distributed along the depth of the cantilever. An increase in the highest $\rho_{GND}$ region just in front of the crack front (marked in red) was also observed throughout Figures \ref{fig:03} g, h and i.

\begin{figure*}[!ht]
\begin{center}
\includegraphics[width=0.95\textwidth]{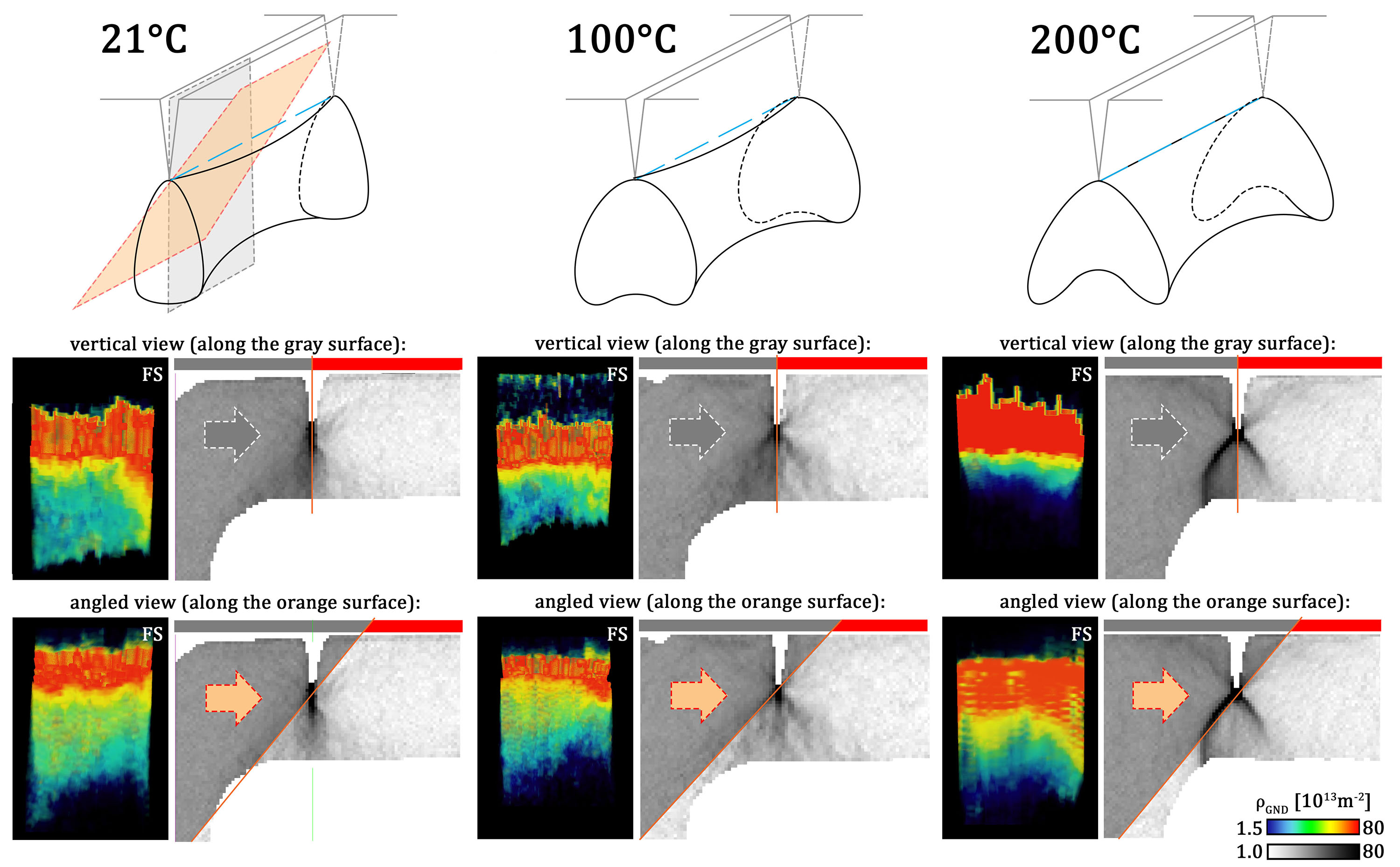} \\
\caption{\label{fig:R1} Evolution of the plastic zone based on the GND density distributions at the three studied temperatures. Vertical (middle row) and angled view (bottom row) show different cross sections of the 3D reconstructions made along the orange line on the grayscale GND density maps. "FS" marks the free surface, while arrows indicate the viewing direction. Sketches of the "dog bone" plastic zone shape in the top row are based on the cross sectional views. Blue dashed line indicates a straight crack front.}
\end{center}
\end{figure*}

However, if different cross-sections are chosen (as presented in Figure \ref{fig:R1}), the presence of this "dog bone" plastic zone can be recognized, but its shape has however changes with increasing temperature. Indeed, the results show a respective increase of GNDs along the main (110) slip line generated at the crack tip. Along with the overall increase of GND density at the crack front, the dislocation pile-up along the slip line decrease dislocation mobility and result in more efficient crack shielding, increasing plastic deformation. These observations are summarized in the sketches in the top row of Figure \ref{fig:R1} to help visualizing how the evolution of the plastic zone takes place at different deformation temperatures.

The shape evolution presented here corresponds well to recent simulation studies on ductile fracture \cite{Noll2019} and experimentally detected on macroscopic crack tension spe\-ci\-mens \cite{Hart2020}. Only a little is known about the 3D shape of the plastic zones in micron-scale fracture specimens, and the current work presents its first characterization attempt. These results show that the “dog bone” shape prevail at every studied temperature conditions, similar to the macro-scale plastic zone distribution.

\section{Discussion}

Although the same displacements were applied during the deformation processes on all cantilevers, a clear difference in crack extension at various temperatures can be observed. From the crack length measurements shown in Figure \ref{fig:R2} it can be concluded that there is slightly longer crack propagation at room temperature in the middle of the sample compared to the edge (about 450 nm). Increasing the deformation temperature to 100$^{\circ}$C this tendency is reduced, where a crack extension in the middle of the cantilever of $\sim$200 nm is measured compared to the edge. The symmetric shape of this extension along the crack tip is also observed. At 200$^{\circ}$C no significant crack propagation was detected. The development of the 3D shapes of the plastic zones clearly shows an effect on the crack extension, where the more pronounced "dog bone" shape of the plastic zone at room temperature is linked to longer crack propagation in the centre of the cantilever.

\subsection{3D GND density evaluation from two different cutting directions}

A sample (S3) with the same dimensions and almost identically deformed at room temperature as the one in Figure \ref{fig:03} a, d and g (S4) was lifted out and milled from the front side of the cantilever. This milling direction is therefore perpendicular to the milling direction of the samples shown in Figure \ref{fig:03}. The advantage of this milling direction is to have the possibility to view the entire volume of the cantilever rather than only the half in case of the side cut milling. 

\begin{figure*}[!ht]
\begin{center}
\includegraphics[width=0.8\textwidth]{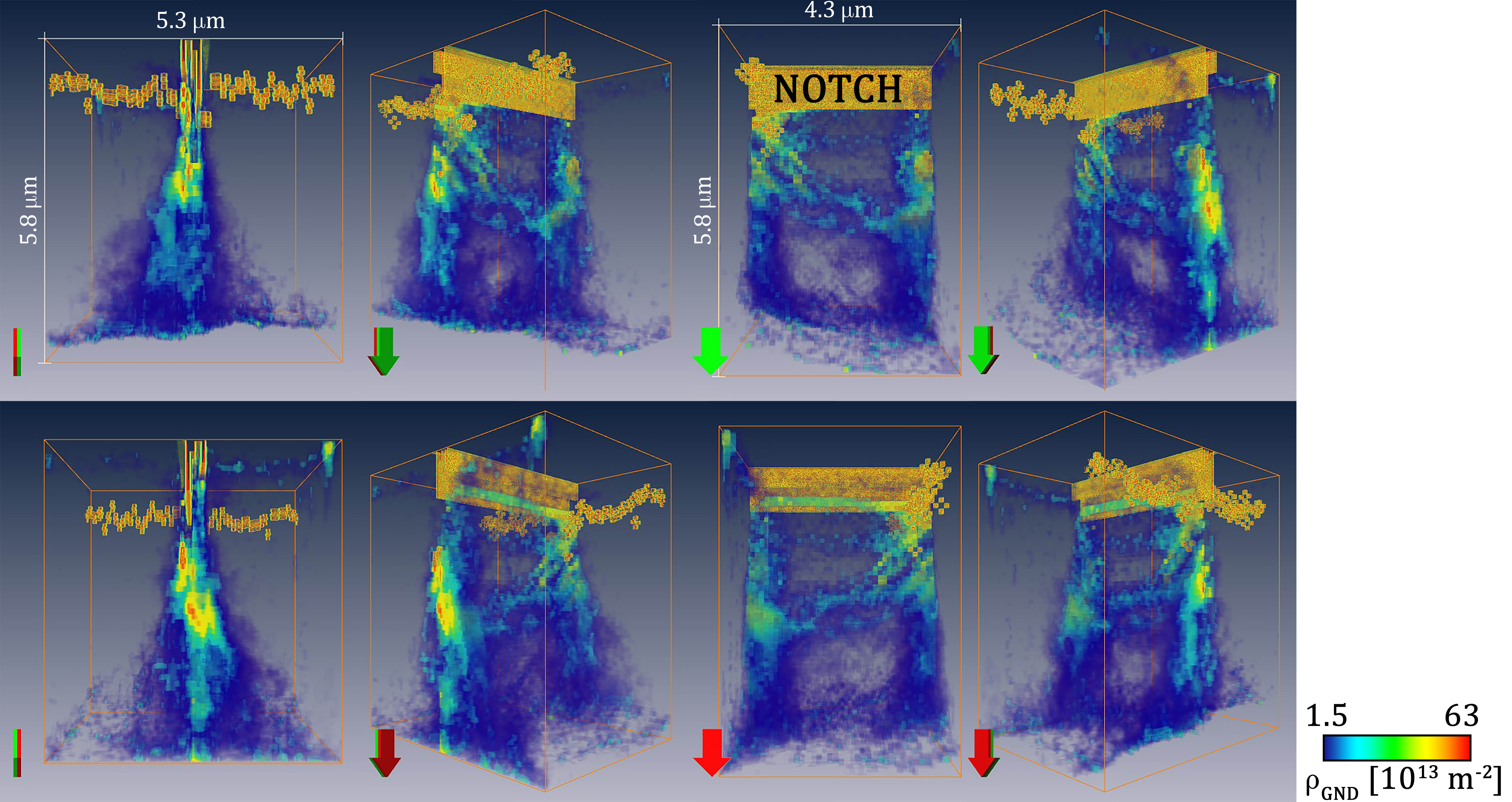} \\
\caption{\label{fig:04} $\rho_{GND}$ values are plotted for the whole sample thickness in the lift-out cantilever (S3). The reconstruction is rotated around for better perception. Rotating green-red arrows in the bottom left corners indicate the orientation of the rendered volume. The position of the crack is indicated by the yellow stripe marked with "NOTCH" on the top.}
\end{center}
\end{figure*}

Figure \ref{fig:04} shows the total $\rho_{GND}$ values  for the whole volume of the lift-out cantilever deformed at room temperature (S4). The cantilever shows a slight tapering in thickness at the bottom perpendicular to the free surface due to FIB milling, that created a slight asymmetry. It is clear from the 3D reconstruction that more GNDs appear close to the free surfaces of the cantilever and in front of the crack tip, making the shape of the plastic zone similar to what is known at larger scale in compact tension specimens (where it is referred to as "dog bone" shape) \cite{Anderson2005}. Therefore, the conventional concept which assumes a constant stress intensity factor along the crack front, a state of plane strain inside the specimen and a state of plane stress at the surface of the specimen prevail at the micron scale.

The depth of the area with high GND density along the free surface is about 1 $\mu$m. One can therefore expect that this is a critical scale for size effect in fracture mechanics, below which the accumulation of dislocations would change around the crack tip. This is confirmed by earlier studies, which shows a change in critical behaviour of both FCC and BCC specimens smaller than 1 $\mu$m \cite{Zhao2019, Abad2016}. This size effect was interpreted as a dislocation starvation phenomenon, where most of the dislocations generated at the crack tip would leave the system immediately through the free surface. In the case of 5 $\mu$m thick test pieces, we are in a mixed behaviour regime, where some dislocations can escape, but there are enough pile-ups near the surface to keep a large amount of GNDs inside the specimen.

\begin{figure}[!ht]
\begin{center}
\includegraphics[width=0.45\textwidth]{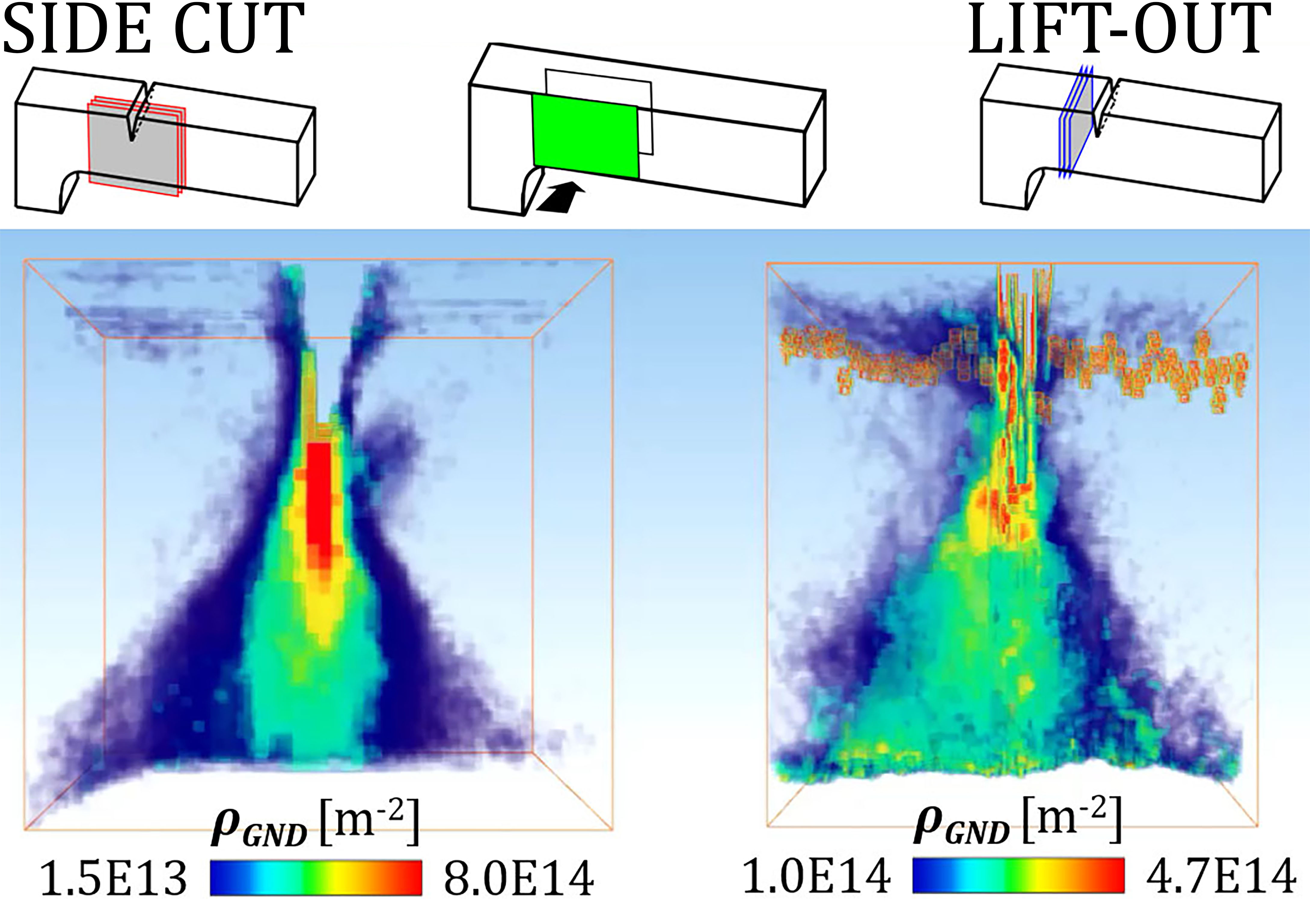} \\
\caption{\label{fig:05} Total GND density values calculated for the similarly deformed cantilevers (left: 21$^\circ$C, S4 - side cutting and right: 21$^\circ$C, S3 - lift-out geometries), viewed from the surface of the sample (green area in the top middle sketch). The left model was created by slicing the cantilever by FIB parallel to its’ side, while the right model was created by slicing the sample perpendicular to the other direction, as shown on the sketches on the top.}
\end{center}
\end{figure}

\begin{figure}[!ht]
\begin{center}
\includegraphics[width=0.45\textwidth]{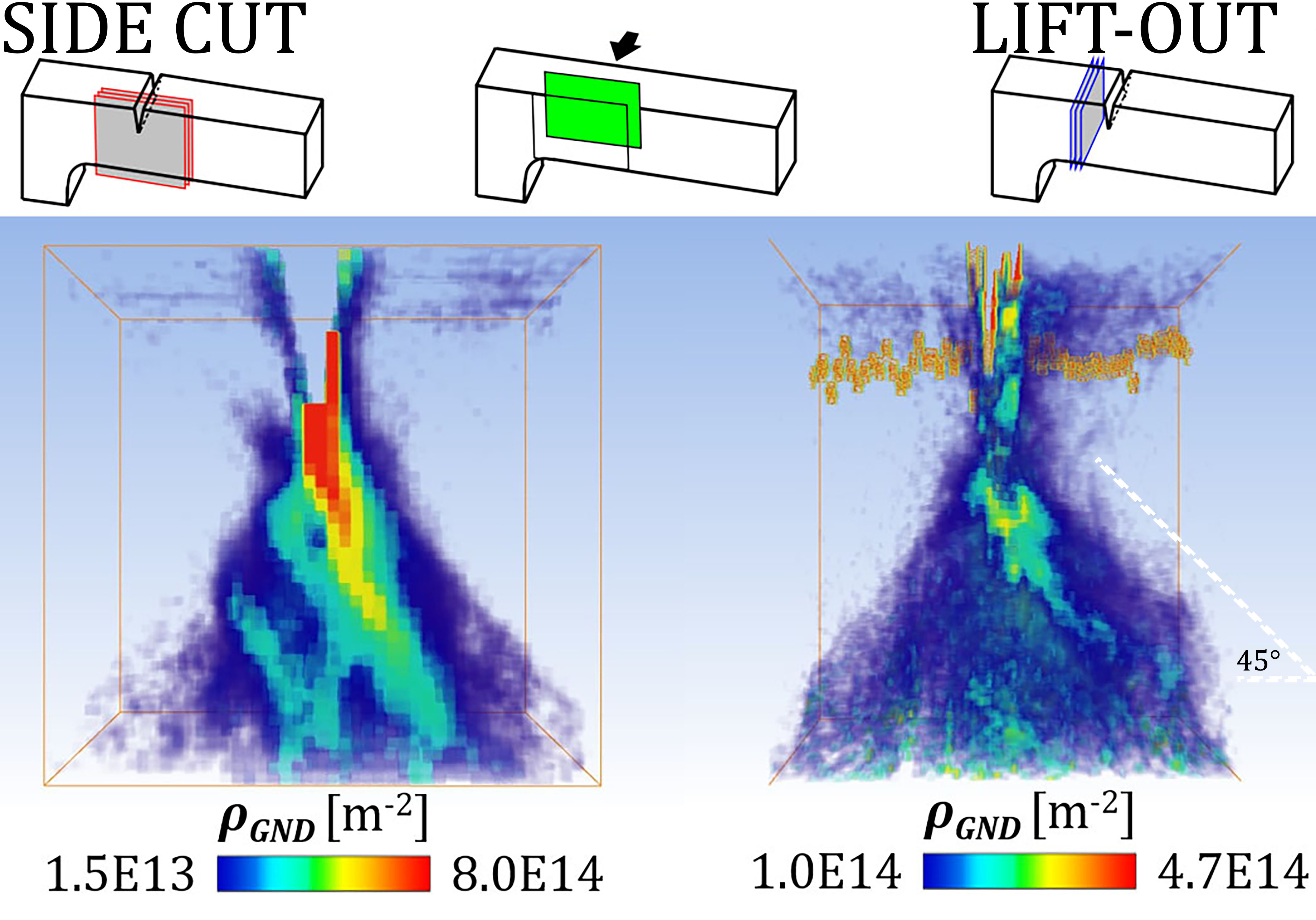} \\
\caption{\label{fig:06} Total GND density values calculated for the similarly deformed cantilevers (left: 21$^\circ$C, S4 - side cutting and right: 21$^\circ$C, S3 - lift-out geometries), viewed from the middle of the sample (green area in the top middle sketch). The left model was created by slicing the cantilever by FIB parallel to its’ side, while the right model was created by slicing the sample perpendicular to the other direction, as shown on the sketches on the top.}
\end{center}
\end{figure}

Total GND density reconstructions of the two cantilevers measured with two different FIB slicing directions (side cutting and lift-out) are compared in Figure \ref{fig:05} and \ref{fig:06}, viewed from the free surface and the middle of the cantilever, respectively. For this comparison, only half of the lift-out reconstruction is shown (the same volume as the other mapped cantilever). Similar features in both reconstructions are detected, namely that the highest concentration of GNDs are located directly in front of the crack tip, and the contour of the plastic zone corresponds well to the (110) type slip planes (mark the $45^{\circ}$ angle in Figure \ref{fig:06}). Interestingly, viewed from the middle of the cantilever in Figure \ref{fig:06}, highest GND concentration seems to be different than what is visible in Figure \ref{fig:05}. It is surprising to see that on the free surface of the cantilever highest GND density was detected straight in front of the crack tip, while in the mid-section a deviation from this primary direction appears and it is angled towards the fixed end of the cantilever. This is an indication that GND distribution along the beam bending axis is not homogeneous. From earlier studies \cite{Ast2018, Ast22018} it is already known that the crack will propagate along the $[0\bar{1}0]$ direction. At such small deformations the (110) slip plane which has the highest Schmid factor (0.48) is activated.

Magnitudes of $\rho_{GND}$ estimated on the lift-out sample show lower values than measured in the side cutting case. To understand the reason behind this phenomenon, the in-depth study of the HR-EBSD calculation is necessary. Lattice distortions that are used to calculate stress/strain tensor components (and GND densities) are not measured equally in the two cases. As already mentioned in section \ref{sec:3D}, only the $\alpha_{i3}$ components can be calculated exactly, while the remaining 6 components are estimated through an optimisation method.  $\alpha_{i3}$ components in the lift-out case correspond to different $\alpha_{ij}$ components in the side cutting case. $\rho_{GND}$ is calculated from all components of the $\alpha_{ij}$ tensor, but if different slicing direction is applied, 3 different $\alpha_{ij}$ components are calculated while others are only estimated. This is an important feature to keep in mind when $\rho_{GND}$ estimation is carried out based on HR-EBSD measurements. Therefore, $\rho_{GND}$ magnitudes should not necessary be the same. Differences in GND density magnitudes or distributions hence originate from detecting different dislocations. This emphasizes the fact that absolute GND density measurement by HR-EBSD need to be taken with care, as it is direction dependent. But comparing different samples with a similar sectioning orientation is still relevant. To better understand what components of the Nye tensor are measured with the two geometries, the Reader is referred to \emph{Appenix B}. If it was possible to do the cross-correlation calculation itself in 3D (which is currently not available by HR-EBSD, only a synchrotron-based Differential Aperture X-ray Laue Micro-diffraction (DAXM) \cite{Guo2020} measurement have been published about this topic), it could give back similar magnitudes and distributions of GND density for the two similarly deformed cantilevers. 3D reconstructions are provided as supplementary videos.

\subsection{Evaluation of the individual components of the Nye tensor}

\begin{figure*}[!ht]
\begin{center}
\includegraphics[width=0.95\textwidth]{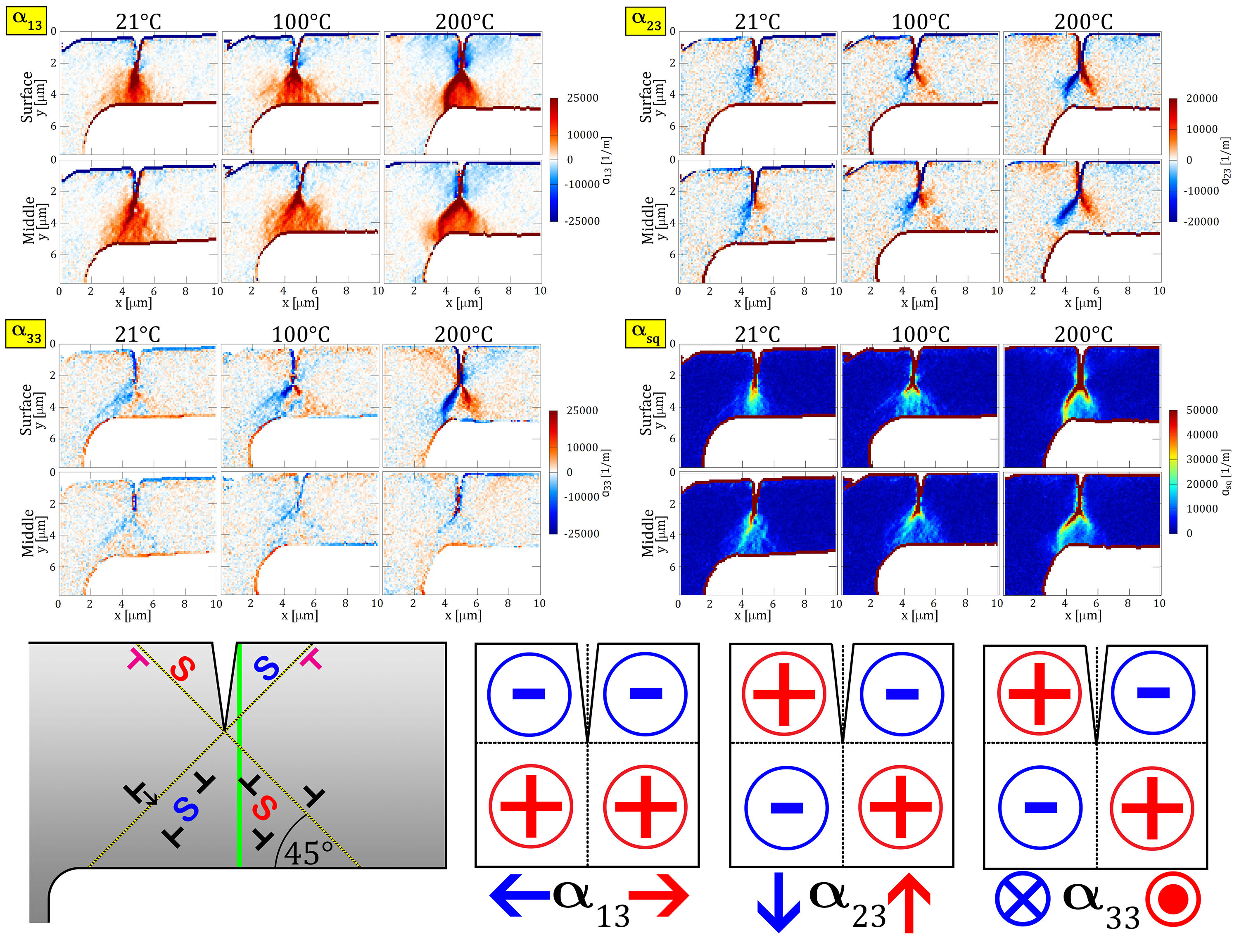} \\
\caption{\label{fig:07} Distribution of $\alpha_{13}$, $\alpha_{23}$, $\alpha_{33}$ and $\alpha_{sq}$ components at three different temperatures close to the surface and in the middle of the cantilever. Bottom row represents a sketch of GND configuration based on the $\alpha_{i3}$ component analysis and Burgers vector directionality. Green vertical line indicates the place from where the lift-out $\alpha_{i3}$ component analysis shown in Fig. \ref{fig:08} was carried out. Arrows indicate the typical Burgers vector direction based on the component analysis. Dislocations coloured in magenta correspond to increased edge type GNDs at higher temperatures due to increased pile-ups in the system.}
\end{center}
\end{figure*}

Usually dislocations in metals do not have pure edge or screw character, but often exhibit a mixture of both. There is however only the possibility to assume pure edge or screw GND components through HR-EBSD and the construction of the Nye tensor. Nevertheless, these components can be individually calculated, giving more insight about local dislocation mechanisms and collective behaviour.

The study of the $\alpha_{i3}$ components were conducted on each cantilever sliced from the side using a section close to the surface (FIB slice No. 4 from the surface) and one map from the middle (FIB slice No. 25 from the surface). $\alpha_{13}$ and $\alpha_{23}$ components in Figure \ref{fig:07} mainly correspond to pure edge type GNDs, while $\alpha_{33}$ mainly represents a pure screw component. The results show, as a general trend for all cantilevers independently of the temperature, that $\alpha_{13}$ and $\alpha_{23}$ components are more homogeneously distributed in the depth of the cantilevers, with no significant difference between the near surface section and the middle of the cantilever. These two components also progressively increase with temperature. However, the $\alpha_{33}$ component shows a clear difference in depth of the cantilever, with less GNDs in the middle of the specimen compared to the near surface section for the three temperatures. One can assume that dislocations are nucleated along the entire crack front (centre and close to the free surface). This is also necessary to accommodate for curvature throughout the specimen thickness during bending. Their type is mixed (screw and edge components, as seen from EBSD results). Since the stress state is more relaxed at the free surfaces than in the centre, dislocations nucleated at the crack front closer to the free surfaces have a higher mobility there, hence we measure higher GND densities in that region. This effect is better visible at room temperature (close to BDT), since at elevated temperatures dislocations have the thermal activation energy to glide more easily and the effects of free surfaces cannot be easly observed any longer.

Based on the GND density analysis, the transition between plain strain (middle of the specimen) to plain stress state at the surface happens in a $\sim 1$ $\mu$m thick region, as discussed earlier. If sample sizes approach this characteristic length the gradient in this transition gets even stronger, that will manifest in size-dependent mechanical effects (the smaller the specimen size, the higher the gradient will be).

The amount of $\alpha_{33}$ GND component close to the free surface is increasing with temperature compared to the $\alpha_{13}$ and $\alpha_{23}$, showing similar trend as in ref. \cite{Ast2018}, where it was also shown that BDT is mainly controlled by the nucleation of screw dislocations in front of the crack tip. As shown in Figure \ref{fig:07}, edge type GNDs are more homogeneously distributed throughout the cross-section of the sample, while screw GNDs exhibit inhomogeneous accumulation and are generally less present in the sample. Screw GNDs clearly tend to be more present towards the free surface of the cantilever than in the center of the cantilever for all deformation conditions. They are therefore the ones that contribute the most to the dog-bone shape of the plastic zone.

Based on the $\alpha_{i3}$ components, the Burgers vector directions can be identified (see also in \emph{Appendix B}). A sketch in the bottom row of Figure \ref{fig:07} reflects only the collective sign of the projected Burgers vector around the crack tip. Earlier studies have already investigated BCC systems, where dislocations emitted from a crack tip were described and crack tip deformation was presented \cite{Ohr1984} and modelled \cite{Tarleton2009, Zavattieri2009}. However, after studying the Burgers vector configuration based on the Nye-tensor components, a conclusion can be made that the GND configuration is not as simple as it was expected earlier \cite{Ast22018}. To our knowledge, the Burgers vector analysis based on HR-EBSD measurement has been performed for the first time to study the directionality of $\underline{b}$. Looking at the tensor component distributions, the direction of the Burgers vector does not only depend on the sign of $\alpha_{ij}$, but also on the differences in magnitudes between different components. 

The stress field around a mode I crack tip in an isotropic medium in plane strain or plain stress state is given in Ref. \cite{rice1967mechanics}. If one then considers two slip planes, one parallel with the $x=y$ plane and one parallel with the $x=-y$ plane, then the resolved shear stresses are everywhere equal on these two slip planes. Based on this, an obvious prediction regarding the activated slip system cannot be made.

In addition, dislocation slip in BCC materials is more complicated than in FCC materials due to the complex core structure of the dislocations. The difficult motion of the screw dislocations involves temperature activated kink-pair formation and wavy slip, where new kinks can form on different slip planes, e.g. \{110\}, \{112\} or \{123\}) \cite{weinberger2013slip}. It's still not understood on which planes dislocations slip and why. Anomalous slip can also occur which means that dislocations due to local stresses can be activated on slip planes with lower resolved shear stresses. Marichal \emph{et al.} \cite{marichal2014origin} showed that these processes are activated through dislocation in\-ter\-ac\-tions, so to understand the plasticity in BCC metals dislocation interactions play a fundamental role \cite{weygand2015multiscale, butler2018mechanisms}. The observations of Fig. \ref{fig:07}, i.e. that Burgers vectors have a direction perpendicular to the line connecting the dislocations and the crack-tip indicate that the detected GNDs are not mainly emitted from the crack tip. It is suggested that they are formed by secondary dislocation sources that are being created by increased dislocation interactions due to large local stresses, pinning, and junction formation.

$\alpha_{sq}$ values show similar concentration in their distribution as the total GND density reconstructions in Figure \ref{fig:03} a, b and c, where the $L_1$ optimisation scheme was used to estimate $\rho_{GND}$ magnitudes. From the $L_1$ optimization method an average GND density in the order of $2 \times 10^{14}$ $1/m^2$ for the dislocation-rich regions and $5 \times 10^{13}$ $1/m^2$ for the middle homogeneous region was estimated.

\begin{figure}[!ht]
\begin{center}
\includegraphics[width=0.49\textwidth]{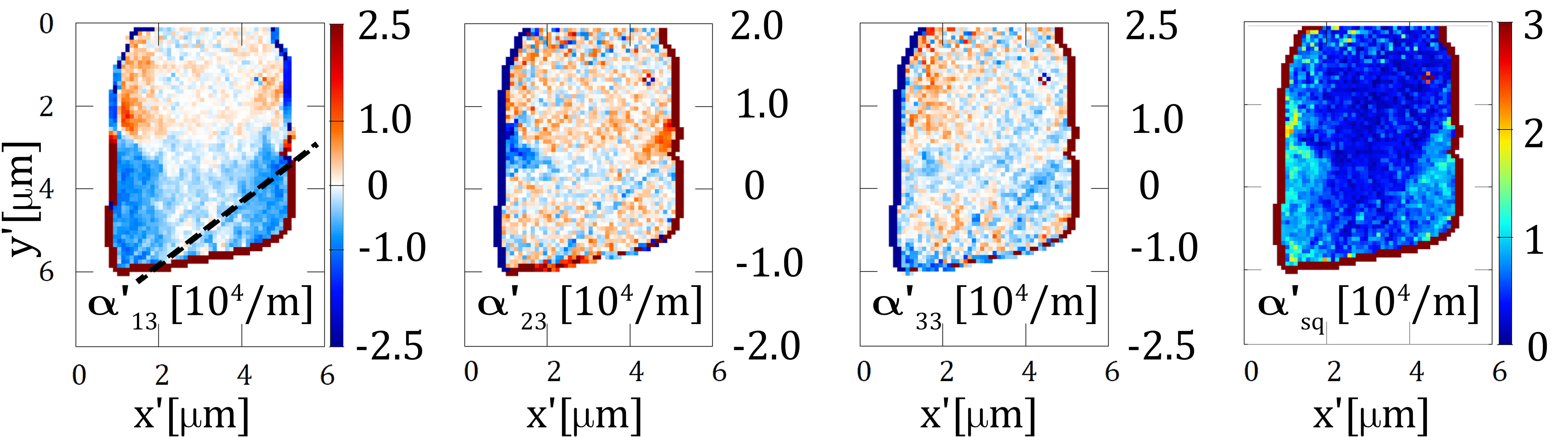} \\
\caption{\label{fig:08} $\alpha_{i3}'$ results on the lift-out sample (S3). Dashed line in the $\alpha_{13}'$ map denotes FIB curtaining effect that has a specific SEM system related angle. This slice was positioned close to the notch, as seen in Figure \ref{fig:07} bottom sketch, marked by a green line. The maps are viewed from the fixed end of the cantilever (also shown as a grey surface in Figure \ref{fig:A1} bottom row sketches).}
\end{center}
\end{figure}

$\alpha_{i3}'$ component analysis was also conducted in the lift-out case (S4) for comparison. One section from the cantilever deformed at room temperature was analysed. Note that the coordinate system is oriented differently in the lift-out and side cut case (see in Fig. \ref{fig:A1}), hence the $\alpha_{i3}'$ denotation. The position of the lift-out slice is shown in Figure \ref{fig:07} bottom row sketch with a green line. This slice was chosen close to the crack, but not directly under it. In Figure \ref{fig:08} the $\alpha_{13}'$ component distribution indicate edge GND pile-ups close to the vertical free surfaces with opposite signs of Burgers vector in front of and behind the crack front ($\alpha_{13}'$). Surprisingly only the middle section ($x'=(2-4)$ $\mu$m) has homogeneous Burgers vector distribution, where "dipole-like" GND arrangement can be observed. Close to the vertical free surfaces on both sides, the edge type GNDs separate by opposite Burgers vector directions ($\alpha_{23}'$). This segregation in Burgers vector directions is only visible until the characteristic depth of 1 $\mu$m below the free surface. On the other hand, $\alpha_{33}'$ component does not show any clear tendency on the direction of the Burgers vector. $\alpha_{sq}'$ values are lower compared to the side cutting case in Figure \ref{fig:07}, that can be described with the difference in Nye tensor calculation already discussed before. Note that the FIB cutting artefact (highlighted with a dotted line in Figure \ref{fig:08} which has a well-defined angle on all maps with $\sim 37^{\circ}$ angle) is introduced during the slicing process, therefore it does not affect the already formed dislocation structure.

It is important to point out that differences in the dislocation structure between the free surface and the interior of the cantilever appears mostly in the $\alpha_{33}$ component (Fig. \ref{fig:07}), therefore the screw component is the most sensitive to the stress gradient. This could be the key difference in the plastic zone shape observed in the 3D reconstruction, and further investigation of the Nye tensor components can provide further insights to complex dislocation structures in metals in the future.

\section{Summary and conclusion}

The 3D shape of the plastic zone in front of crack tips has been successfully characterized by 3D HR-EBSD. As model material and geometry, single crystal W microcantilevers with a notch were deformed at room temperature, 100$^{\circ}$C and 200$^{\circ}$C at various deformation rates. These deformation conditions correspond respectively close to a brittle-ductile regime towards a progressive fully plastic regime at the highest temperature. 

For the deformation at room temperature, the distribution of GNDs in front of the crack tip shows a clear "dog bone" shape, with more GNDs towards the free surface of the cantilever and less GNDs in the centre of the specimen. This is comparable to what is observed in compact tension specimen at the macro-scale. At higher temperatures, in a more extended plastic regime this "dog bone" plastic zone still appears but its shape evolves with higher temperature, increasing the overall dislocation density at the crack tip, which inhibits crack propagation. This behaviour is supported by earlier studies in macroscopic specimens and simulations. The separate analysis of the $\alpha_{i3}$ components of the Nye tensor shows for the first time a pronounced trend for more nucleation of pure screw GNDs at high temperature. Screw GNDs are also the one responsible for the inhomogeneous distribution of GNDs between the centre and the free surface of the cantilever. 

This study reveals the importance of the free surface during micro-mechanical testing and fracture mechanics at small scales. This effect on the mechanism of deformation can only be understood if we are able to characterize in 3 dimensions the distribution of dislocations in deformed materials. Results presented here can be used to compare experiments with simulations. The development of novel computational methods should also focus on how to take into account this free surface effect.

\section*{APPENDIX A-- Fracture toughness calculation}

The geometry factor $f$, which is a function of the crack length and the sample height, was calculated by finite element (FE) simulations for the investigated geometry \cite{Ast2018}:

\begin{eqnarray}\label{eq:03}
\begin{aligned}
f\left( \frac{a}{W}\right) = {} & 1.24 + 28.08 \left( \frac{a}{W}\right)
-59.44\left( \frac{a}{W}\right)^2 \\ &
+ 90.40\left( \frac{a}{W}\right)^3
\end{aligned}
\end{eqnarray}

Elastic-plastic fracture mechanics (EPFM) were used to determine the fracture toughness characterized by the onset of stable crack growth, which is denominated as "fracture initiation toughness" in the following. Continuous J-Integrals composed of an elastic and a plastic part ($J_{el}$ and $J_{pl}$, respectively) were determined for this purpose. The technique, which utilizes a slightly modified approach of the one described in the ASTM standard E1820 \cite{ASTM1820} for macroscopic samples, is explained in more detail in a previous study on microcantilevers \cite{Ast2016}:

\begin{eqnarray}\label{eq:04}
\begin{aligned}
J_{(i)} = {} & J_{el, (i)} + J_{pl, (i)} = \frac{(K_{Iq,(i)})^2(1-\nu^2)}{E} \\&
+\left[ J_{pl, (i-1)} + \frac{\eta(A_{pl, (i)} - A_{pl, (i-1)})}{B(W-a_{(i-1)})} \right] \\&
\times \left[ 1-\frac{a_{(i)}-a_{(i-1)}}{W-a_{(i-1)}} \right] ,
\end{aligned}
\end{eqnarray}
where $\nu$ is the Poisson's ratio ($\nu^W=0.28$), and $E$ is the Young's modulus ($E^{21^{\circ}C}=411$ GPa, $E^{100^{\circ}C}=427$ GPa and $E^{200^{\circ}C}=401$ GPa \cite{Varshni1970}). $K_{Iq,(i)}$ is determined for each oscillation cycle $(i)$ by adapting the previously shown Eq. (\ref{eq:02}) from the ASTM standard E399 \cite{ASTM399}:

\begin{eqnarray}\label{eq:02mod}
\begin{aligned}
K_{Iq,(i)}=\frac{F_{q,(i)}L}{BW^{3/2}} \times f\left( \frac{a_(i)}{W}\right).
\end{aligned}
\end{eqnarray}

\begin{figure*}[!ht]
\begin{center}
\includegraphics[width=0.7\textwidth]{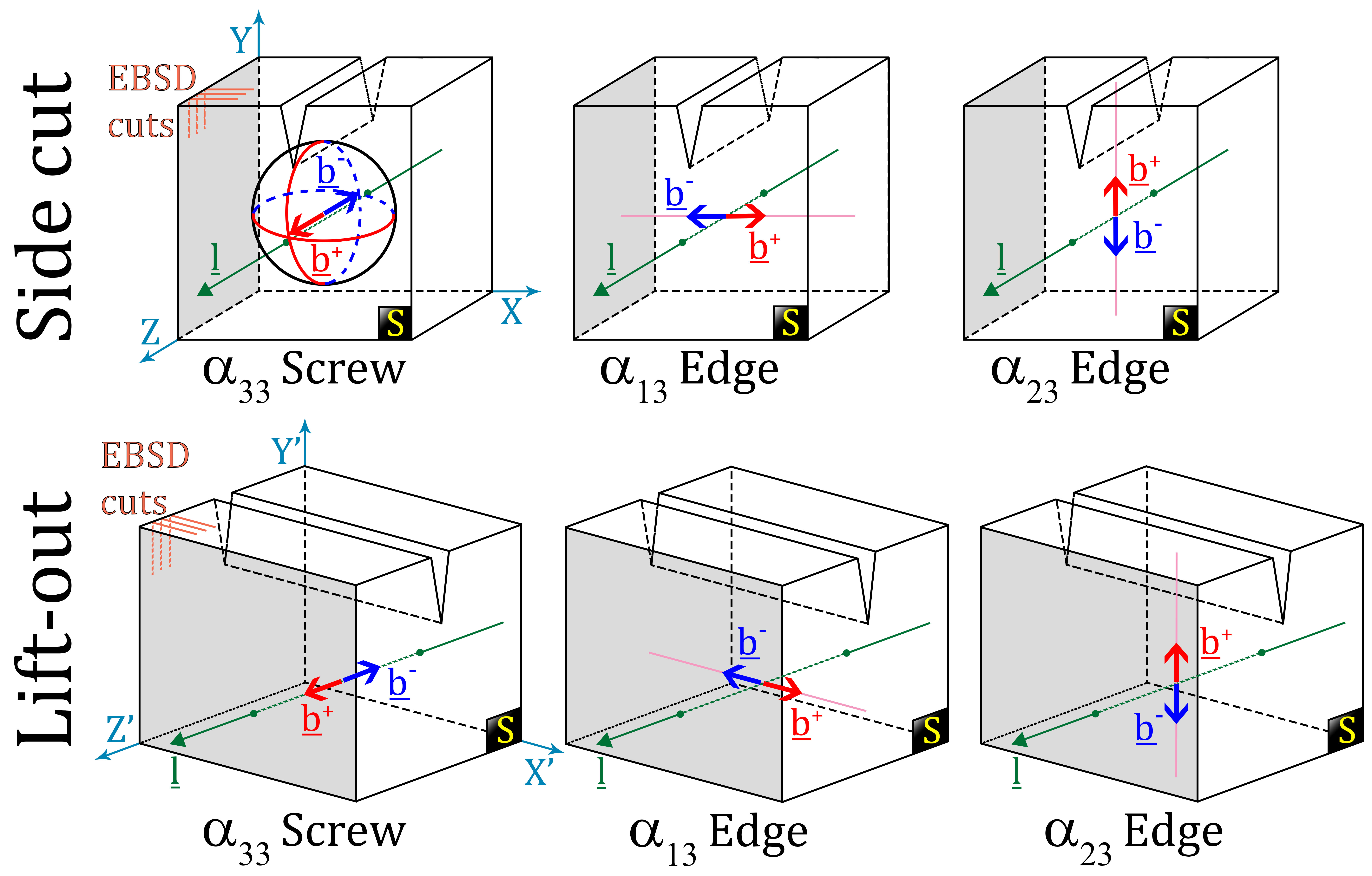} \\
\caption{\label{fig:A1} Sketches of the detectable dislocations in the two measurement cases of side cutting and lift-out geometry. The coordinate systems ($XYZ$ and $X' Y' Z'$) are linked to the EBSD measurement geometry. The sign \colorbox{black!90}{\textcolor{yellow}{S}} indicates the outer side of the cantilever close to the edge of the bulk sample, while the gray surface highlights the side of the cantilever's base. Burgers vectors of all directions represented by a sphere are detected as a projection onto the three main axes.}

\end{center}
\end{figure*}

Here, the deviation from the initial linear loading re\-gime by means of the 95 \% construction line is used to calculate the critical force $F_q$. For the calculation of the plastic part of the J-Integral $J_{pl}$, the area underneath the force-displacement curve $A_{pl}$ is determined. The constant $\eta=2$ is set for the bending geometry. From the obtained R-curves critical J-Integrals $J_q$ are determined according to different criteria for the onset of stable crack propagation. Using the following equation, the conditional fracture toughness is calculated as:

\begin{eqnarray}\label{eq:02final2}
\begin{aligned}
K_{Iq,J}=\sqrt{\frac{J_{q}E}{(1-\nu^2)}}.
\end{aligned}
\end{eqnarray}

\section*{APPENDIX B-- Understanding the measured Nye tensor components by HR-EBSD}

In this article we define the Burgers vector ($\underline{b}$) in the RHFS convention (right hand Burgers loop around the dislocation, and the Burgers vector points from the finishing to the starting point of the loop).

The second rank tensor $\alpha_{ij}$ was introduced by Nye in 1953 \cite{Nye1953} in the form of

\begin{eqnarray}\label{eq:05}
B_i = \alpha_{ij} l_j, \quad (i,j=1,2,3),
\end{eqnarray}
where the $\alpha_{ij}$ geometrically necessary dislocation density tensor coefficients relate the two vectors $B_i$ Burgers vector and $l_j$ line vector to specify the "state of dislocation" in the area. This tensor includes screw type dislocations ($\alpha_{ii}$), where $\underline{b} \parallel \underline{l}$ and edge type dislocations ($\alpha_{ij}, i \neq j$), where $\underline{b} \perp \underline{l}$.

In case of HR-EBSD, as backscattered electrons only originate from the top couple of tens of nanometers of the sample, we can only calculate three components of the Nye tensor from 2D experimental data as

\begin{eqnarray}\label{eq:06}
\alpha_{i3} = \partial_1 \beta_{i2} -  \partial_2 \beta_{i1} \quad i=1,2,3,
\end{eqnarray}
where $\beta_{ij}$ are the components of the deformation gradient tensor \cite{Wilkinson2009}. HR-EBSD analysis by cross-correlation calculation can determine the $\beta_{ij}$ components.

When 3-dimensional sectioning is applied from the side cutting and lift-out case, the measurement coordinate systems in which we define these components differ. In Figure \ref{fig:A1} the two geometries are shown along with the dislocations that we can detect through the $\alpha_{i3}$ components.

\begin{figure}[!ht]
\begin{center}
\includegraphics[width=0.45\textwidth]{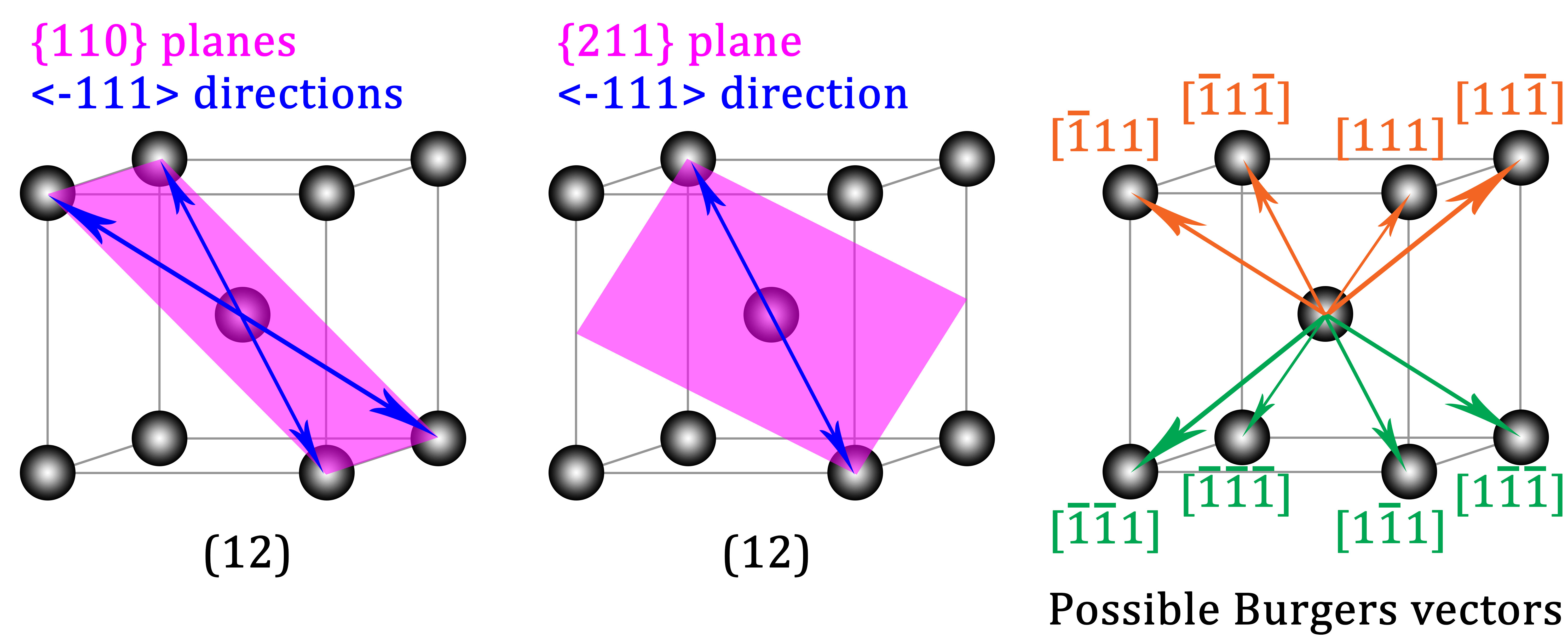} \\
\caption{\label{fig:A2} Main slip systems and Burgers vectors in tungsten. Both slip systems can be activated in W with 12 possible configurations each.}
\end{center}
\end{figure}

Usually dislocations have mixed type and their Burgers and line vectors does not align with the measurement coordinate system. It is important to note that discrete dislocations cannot be detected by HR-EBSD (only lattice rotations), and the same GND can be detected through different $\alpha_{ij}$ components. Therefore, Figure \ref{fig:A1} shows only the ideal case, but it does not mean that we have purely these dislocations. In reality, it is more like the mixture of these components, and as we measure only the projection of the Burgers vector, the differentiation between types is never straightforward. In Figure \ref{fig:A2} possible BCC slip systems have been shown. In W, mainly the {110} and {211} types (24 possible configurations) can be activated (in total 48 types).

As the bulk sample was oriented towards the three main axes of the unit cell, it is easy to see that different projections of the same GND with a Burgers vector $1/2<111>$ type can contribute to more than one Nye tensor component.

\section*{Acknowledgement}

The Authors thank Prof. Karsten Durst and Prof. Istv\'{a}n Groma for many fruitful discussions on small-scale fracture behaviour and dislocation density analysis.
SzK was supported by the EMPA\-POST\-DOCS-II programme, that has received funding from the European Union's Horizon 2020 research and innovation programme under the Marie Skło\-dow\-ska-Curie grant agreement number 754364. PDI was supported by the ELTE Institutional Excellence Program (1783-3/2018/FEKUTSRAT) supported by the Hungarian Ministry of Human Capacities and by the National Research, Development and Innovation Fund of Hungary (contract number: NKFIH-K-119561). The Authors are grateful for the support of Dr. Graham Meaden from BLG Vantage on the debugging process of the Nye-tensor component analysis.

\bibliography{mybibfile}

\end{document}